\documentclass[doublecol]{epl2} 
\usepackage{amsmath,amscd,amssymb,epsfig}
\usepackage{graphicx}

\title{Markovian embedding of fractional superdiffusion}

\author{P. Siegle \and I. Goychuk \and P. H\"anggi}
\shortauthor{P. Siegle \etal}

\institute{                    
   Institute of Physics, University of Augsburg - Universit\"atsstra\ss e 1, D-86135 Augsburg, Germany
}

\abstract{ 
The Fractional Langevin Equation (FLE) describes a 
non-Markovian  Generalized Brownian Motion  with long time persistence
(superdiffusion), or anti-persistence (subdiffusion) of both
velocity-velocity correlations,  and position increments. It  presents
a case of the Generalized Langevin Equation (GLE) with a singular
power law memory kernel. We propose and numerically realize a 
numerically efficient and reliable  Markovian embedding of this
superdiffusive GLE, which accurately approximates the FLE over many, about $r=N\lg b-2$, 
time decades, where $N$ denotes the number of exponentials used to approximate
the power law kernel, and $b>1$ is a scaling parameter for the hierarchy of relaxation
constants leading to this power law.
Besides its relation to the
FLE, our approach presents an independent and very flexible route to
model anomalous diffusion.
Studying such a
superdiffusion in tilted washboard potentials, we demonstrate
the phenomenon of transient hyperdiffusion which emerges due to
transient kinetic heating effects.}

\begin{document}

\maketitle

\section{Introduction}

Anomalous diffusion and transport processes possess a rich variety of
applications spanning many different research fields from plasma
physics and nonlinear dynamical systems to condensed matter physics,
biophysics, epidemiology, and even quantitative finance 
\cite{MetzKlaf,Hughes,Bouchaud}.  There are
several very different theoretical approaches to describe  anomalous
diffusion, from continuous time random walks (CTRW) \cite{Scher} including Levy
flights and Levy walks \cite{MetzKlaf,Hughes} 
to the Generalized Langevin Equation (GLE) 
\cite{Kubo,Zwanzig,HTB90,WeissBook,Coffey,Wang,Mainardi,Lutz,Kupf,GH07,Lu,G09,
Deng,PRE,PRL,Jeon}.
CTRW-based anomalous diffusion  involves such unusual concepts as
fractal time and subordination to a random clock which does not
possess a finite mean period. In the continuous space limit it is often
associated  with Fractional Fokker-Planck Equations (FFPEs) \cite{MetzKlaf}.
The corresponding Langevin equations are local in {\it random} time 
\cite{Langevin} and describe a
doubly random process with infinite memory. 
These latter Langevin equations 
should not be confused \cite{GH07} with the Fractional
Langevin Equations (FLEs) \cite{Coffey,Mainardi,Lutz,Kupf,GH07,Deng,Jeon}.
Random time clocks without mean period entail the remarkable phenomenon of weak
ergodicity breaking \cite{new1,new2}. The position, or  velocity increments in this
approach are independent in all basic models. Ergodicity can also be broken in other
approaches, e.g. in nonlinear Brownian motion \cite{LutzPRL}. 
Furthermore,  
GLEs including FLEs
present a quite different approach incorporating
long-time correlations or anti-correlations of the position increments,
similar to the fractional Brownian motion (fBm) devised by Mandelbrot and van
Ness \cite{Mandelbrot} which can also be derived from a FLE \cite{GH07}. 
This is a benchmark feature \cite{G09,Jeon} of the latter approach. Another one is that this approach 
does not rely on the concept of random time with divergent
mean clock period \cite{G09} and is almost always ergodic,
except for ballistic GLE diffusion \cite{Wang,G09,Deng,PRE,PRL}. 

Generally such non-Markovian processes are not characterized completely by a master
equation for conditional probabilities, or  Fokker-Planck equations \cite{GTH80}.
However, the corresponding Langevin equations specify the stochastic
process completely, containing all the information on trajectories. 
Phenomenologically, the GLE
\begin{equation}\label{gle}
m\ddot x+\int_{-\infty}^{t}\eta(t-t')\dot x(t')dt'+
\frac{\partial V(x,t)}{\partial x}
=\zeta(t) \;,
\end{equation}
formally presents a Newtonian equation
of motion for a particle with mass $m$ and coordinate $x$ (we consider a
one-dimensional model), subjected apart from a regular force, 
$f(x,t)=-\partial V(x,t)/\partial x$, to a zero-mean 
stochastic force $\zeta(t)$
(which adds energy to the Brownian particle) and a non-local in
time  frictional, or dissipative force (which takes off energy
from the Brownian particle). Both processes are balanced at
thermal equilibrium, i.e. Brownian motion never ceases and obeys the
fluctuation-dissipation theorem, FDT. In (\ref{gle}), 
the dissipative force is assumed  to have 
the form of a linear velocity-dependent friction with its memory 
characterized by the integral kernel
$\eta(t)$. The FDT is obeyed, when the noise is Gaussian \cite{Reimann} 
and the memory kernel and the noise autocorrelation function are related by
the fluctuation-dissipation relation (FDR)\cite{Kubo} reading
\begin{equation}\label{fdt} \langle
\zeta(t)\zeta(t')\rangle=k_BT\eta(|t-t'|) \;
\end{equation}
with $T$ being the environmental temperature. 

Apart from this phenomenological justification, 
the GLE can be derived
microscopically from a Hamiltonian model involving coupling of the
diffusing particle to a thermal bath of harmonic oscillators obeying
initially canonical distribution at temperature  $T$ \cite{Zwanzig,HTB90,WeissBook}. 
Anomalous diffusion can be related 
to a power-law memory kernel  
\begin{equation}\label{reg} 
\eta(t)=\frac{|\sin(\pi\alpha/2)|}{\pi/2}\Gamma(\alpha)\eta_{\alpha}
{\rm Re}(it+1/\omega_c)^{-\alpha} \;
\end{equation}
which we write with a short-time cutoff $1/\omega_c$ corresponding
to the largest frequency of the bath oscillators $\omega_c$, $\alpha>0$, 
and $\Gamma(x)$
is the gamma-function. By eq. (\ref{fdt}), for $0<\alpha<2$ 
such $\eta(t)$ corresponds in the singular limit $\omega_c\to\infty$
to a fractional Gaussian noise $\zeta(t)$ 
with the Hurst exponent $H=1-\alpha/2$ \cite{Mandelbrot}. 
Then the solution of the GLE yields a fractional
Brownian motion in the limit $m\to 0$ \cite{GH07}. 
In the case of free diffusion [$f(x,t)=0$] and for $0<\alpha<2$ independently 
of $\omega_c$ the noise-averaged position variance $\sigma^2(t)=\langle\Delta
x^2(t)\rangle=\langle [x(t)-\langle x(t)\rangle]^2\rangle$
grows with time {\it asymptotically}
as  $\sigma^2(t)\sim 2k_BTt^\alpha/[\eta_\alpha\Gamma(1+\alpha)]$ \cite{WeissBook}.
This corresponds to
sub-diffusion in the case $0<\alpha<1$
(sub-linear growth, sub-Ohmic thermal bath), normal diffusion for $\alpha=1$ 
(linear growth, Ohmic bath), and superdiffusion for $1<\alpha<2$ 
(super-linear growth, super-Ohmic bath). In terms of the integral
frictional strength,  $\tilde\eta(0)=\int_0^{\infty}\eta(t')dt'$, these behaviors
are intuitively clear:  subdiffusion corresponds to $\tilde\eta(0)\to\infty$,
normal diffusion to $\tilde\eta(0)=const$ and superdiffusion to $\tilde\eta(0)\to 0$.
For $\alpha>2$, the free diffusion is always ballistic, $\sigma^2(t)\sim t^2$,
and nonergodic because the velocity autocorrelation function (VACF) does not
decay to zero.
Therefore, the emergence of hyperdiffusion $\sigma^2(t)\sim t^\lambda$, with $\lambda >2$
within GLE model is rather surprising. All these results can be easily
obtained from a general expression for the stationary VACF,
found for an arbitrary memory kernel first by Kubo \cite{Kubo} 
and reading
$\tilde{K}_v(s)=k_BT/[ms+\tilde \eta(s)]$ in the Laplace space,
 by taking
into account that the position variance is the twice-integrated VACF. 
One has to remark at this point, that the Laplace-transformed $\eta(t)$ is
$\tilde\eta(s)=\eta_{\alpha}s^{\alpha-1}$ in this model in 
the limit $\omega_c\to\infty$.
However, the inverse Laplace
transform for $\tilde\eta(s)=\eta_{\alpha}s^{\alpha-1}$ only exists for
$\alpha\le 1$, i.e. for normal and subdiffusive friction. For $1<\alpha<2$
and  a finite $\omega_c$, the kernel $\eta(t)$ in eq. (\ref{reg}) starts from
a positive part and then becomes negative so that its total integral is zero,
independently of $\omega_c$. When the cutoff frequency tends to infinity,
the memory kernel becomes singular, starting from a positive singularity and being
negative otherwise. The frictional term can be recast in this limit with the
help of a fractional Riemann-Liouville derivative, see below.

For example, a spherical particle of radius $R$ moving with velocity $v(t)=\dot x(t)$
in an incompressible liquid of kinematic viscosity $\mu$ and density $\rho$ experiences a hydrodynamic
force \cite{Landau}
\begin{eqnarray}
 F_v(t)&=&-2\pi\rho R^3\Big(\frac{1}{3}\ddot x+\frac{3\mu}{R^2}\dot x\nonumber\\
 && +\frac{3}{R}
\sqrt{\frac{\mu}{\pi}}\int_{-\infty}^t
\frac{\ddot x(\tau)}{\sqrt{t-\tau}}d\tau  \Big) \;.
\end{eqnarray}
This classical result due to Boussinesq and Basset \cite{Mainardi} 
generalizes the well-known  by Stokes.
The first term yields
a mass renormalization of the Brownian particle $m\to m+\Delta m$ with 
$\Delta m=2\pi \rho R^3/3$, which is assumed to be implicitly done. 
It is present also in the absence of dissipation, i.e. 
for $\mu\to 0$. The second term corresponds to
the Stokes friction, and the third term 
is due to a finite relaxation time $\tau_r=R^2/\mu$ of the 
disturbed velocity field of the liquid. It reflects 
hydrodynamic memory.  An interesting
mathematical interpretation of this term can be given within the formalism of 
fractional derivatives 
\cite{MetzKlaf,Mainardi}. 
Namely,  using the definition of the Riemann-Liouville fractional derivative, 
$\sideset{_{t_0}}{_t}{\mathop{\hat D}^{\gamma}}f(t):=\frac{1}{\Gamma(1-\gamma)}
\frac{d}{dt}\int_{t_0}^t dt' f(t')/(t-t')^\gamma$, $0<\gamma<1$, 
acting on some test function $f(t)$,
it can be recast in the form \cite{Mainardi}
\begin{eqnarray}\label{model1}
 F_{ad}(t)= -\eta_{\alpha}\sideset{_{-\infty}}{_t}{\mathop{\hat D}^{\alpha-1}} \dot x(t)
\end{eqnarray}
with  $\eta_{\alpha}=\eta_1R/\sqrt{\mu}$ and $\alpha=3/2$, where 
$\eta_1=6\pi R\mu\rho$ is the Stokes friction coefficient. The corresponding GLE
was termed FLE \cite{Mainardi}. 
Independently of this interpretation it is known \cite{KuboBook} to 
yield the famous power-law decay of the VACF, 
which was revealed in molecular
dynamic simulations by Alder and Wainwright \cite{Alder}. 

The presence 
of a normal Stokes friction term makes the corresponding diffusion asymptotically
normal. However, anomalous superdiffusive motion can emerge on a
transient time scale $t<\tau_r=R^2/\mu$ for light particles \cite{Mainardi}. 
If to neglect {\it ad hoc} the Stokes term and to set
$\dot x(t)=0$ for $t<t_0=0$ (i.e. the particle starts to move at $t=0$)
the corresponding superdiffusive FLE 
reads \cite{Mainardi,Coffey},   
\begin{equation}\label{fle}
m\ddot x+ \eta_{\alpha}\sideset{_{0}}{_t}{\mathop{\hat D}^{\alpha-1}} \dot x(t)+
\frac{\partial V(x,t)}{\partial x}
=\zeta(t) \;
\end{equation} 
with $\alpha=1.5$. This FLE 
serves as one of the basic models for superdiffusion with $1<\alpha<2$.
It corresponds to a GLE with a singular memory kernel, which in mathematical sense 
is a generalized function. 
As discussed above, this kernel starts from a 
positive singularity at $t=0$ and then is negative, decaying to zero in accordance
with a power law $t^{-\alpha}$, so that its total integral is zero.

The presence of a nonlinear time-dependent force $f(x,t)=-\partial
V(x,t)/\partial x$ modifies this well-established picture considerably.
Since general
analytical results are then scarce and most likely generally nonexistent, 
the reliability of numerical simulations
is a key issue. In particular, 
numerically tractable models can be obtained by approximating the
given power law memory kernel by a finite sum of exponentials, 
see in \cite{Kupf,G09,PRE}.
By means of this approximation it is possible to represent the non-Markovian 
dynamics in the $(x,v)$ plane as a projection of a fully Markovian dynamics
in a hyperspace of dimension $D=N+2$, where $N$ is the number of
exponentials in the approximation \cite{Kupf,G09,PRE}. Then the central point is that one can
propagate the corresponding Markovian dynamics locally in time by
very reliable algorithms with a well controlled numerical precision and by increasing
$N$ one can approximate the FLE dynamics ever better. Surprisingly one finds that
the practical embedding dimension $D$ need not be large to achieve an excellent
approximation within statistical errors of stochastic simulations. 
Moreover, this approach can be used
independently of the FLE with some advantages:
(i) The stochastic propagation 
is local in time
and can readily be continued beyond end point. There is no need to generate
a realization of long-correlated noise $\zeta(t)$ for the {\it whole} time span
of simulation {\it fixed} in advance (without a possibility to continue) and
solving the intregro-differential equation numerically for any such noise
realization. This dramatically saves computer memory and enables extremely long
simulations with appreciably small statistical errors. (ii) The corresponding 
multi-dimensional Fokker-Planck equation is known explicitly for arbitrary $f(x,t)$ which can 
be used to develop an analytical theory for nonlinear dynamics. 

\section{Model}

\begin{figure}
  \onefigure[width=.45\textwidth]{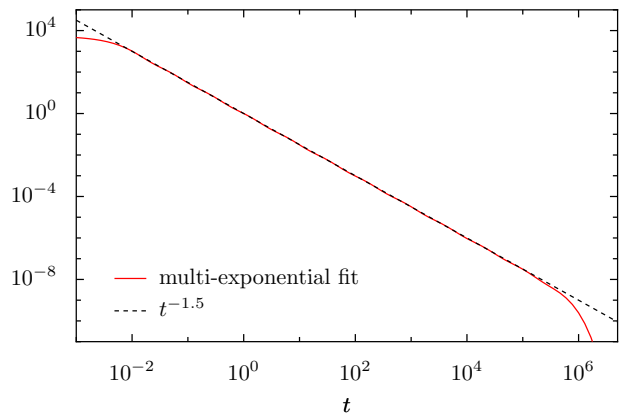}
  \caption{Approximation of the power law behavior of the friction kernel
  by a sum of exponentials $C_\alpha(b)\sum^N_{i=1}
  \left(\nu_0/b^i\right)^\alpha\exp\left(-\nu_0t/b^i\right)$
  [see eq. (\ref{fk})] with the parameters: $N=13$, $\nu_0=10^3$, 
  $b=5$, and $C_\alpha(b)=1.78167$.}
  \label{fig1}
\end{figure}

We generalize the modeling of viscoelastic subdiffusion 
in ref.\cite{G09} to superdiffusive
case, $1<\alpha<2$, and approximate the superdiffusive memory kernel as    
\begin{align}
  \eta(t)=\frac{\eta_\alpha C_\alpha(b)}{|\Gamma(1-\alpha)|}
  \Bigg[&2\sum^N_{i=1}\left(\frac{\nu_0}{b^i}\right)^{\alpha-1}\delta(t)\nonumber\\
  &-\sum^N_{i=1}\left(\frac{\nu_0}{b^i}\right)^\alpha
  \exp\left(-\frac{\nu_0}{b^i}t\right)\Bigg]\;.
  \label{fk}
\end{align}
The first singular term mimics the positive singularity in the memory kernel of the
FLE (\ref{fle}). The sum of exponentials obeys a fractal scaling  
with negative weights and approximates the
power law decay \cite{Hughes} 
of this memory kernel; i.e., $\eta(t)\propto - t^{-\alpha}$, for $t>0$,
so that $\tilde \eta(0)=0$. Choosing $\nu_0=\omega_c$, 
the power law regime extends in this approximation from
a short time (high-frequency) cutoff, $\tau_l=\omega_c^{-1}$, to a large time 
(small frequency) cutoff, $\tau_h=\tau_l b^N$, where $b$ is a scaling dilation
parameter. Such a fit is known to exhibit logarithmic oscillations superimposed on
the power law \cite{Hughes}. 
Their amplitude is, however, small and can be controlled by the 
choice of $b$. By adjusting $b$ and $N$ for a given $\alpha$ one can 
approximate $t^{-\alpha}$ over about $r=N\lg b-2$ time decades between
two time cutoffs, which are always physically present, beyond FLE modeling. 
Fig. \ref{fig1} illustrates the quality of the approximation of $t^{-1.5}$
with the parameters:  $\nu_0=10^3$, $b=5$, $N=13$, and $C_\alpha(b)=1.78167$.
One can detect a good agreement over about $r\approx 7$ decades in time. 
Clearly, with decreasing $b$ and increasing $N$, one can further improve and 
control the quality
of the approximation \cite{G09} which
should be consistent with statistical errors of Monte Carlo simulations
to avoid unnecessary numerical load.

Next, we introduce $N$ auxiliary variables $u_i$ and the corresponding 
multi-dimensional Markovian dynamics in the hyperspace of dimension $D=N+2$, 
\begin{align}\label{emb}
\dot x(t)=&v(t)\nonumber \\
 m\dot v(t)=&-V^\prime(x,t)-\sum^N_{i=1}u_i(t)-\eta_0
 v(t)+\sqrt{2k_BT\eta_0}\xi_0(t) \nonumber \\
 \dot u_i(t)=&-\eta_i v(t)-\nu_iu_i(t)+\sqrt{2k_BT\eta_i\nu_i}\xi_i(t) ,
\end{align}
where $\nu_i=\nu_0/b^i$,
$\eta_i= C_\alpha(b)\eta_\alpha\nu_i^\alpha/|\Gamma(1-\alpha)|$,
for $i=1,...,N$, and $\eta_0=\sum_{i=1}^N\eta_i/\nu_i$.
Furthermore, $\xi_0(t)$ and the $\xi_i(t)$ are $N+1$ 
delta-correlated white Gaussian noise sources of zero-mean and unit intensity.
$N$ of them are totally uncorrelated, 
$\langle\xi_i(t)\xi_j(t^\prime)\rangle=\delta_{ij}\delta(t-t^\prime)$, for
$i,j=1,...,N$, and for $i=0,j=0$. 
However, the noise $\xi_0(t)$ is chosen as a weighted, normalized sum of the other
{\it independent} noises, 
\begin{eqnarray}\label{corr}
\xi_0(t)=\sum_{i=1}^N\sqrt{\frac{\eta_i}{\nu_i\eta_0}}\xi_i(t) \;.
\end{eqnarray}

In the limiting case $N=1$, our present model yields the minimal 3-dimensional
embedding of ballistic GLE superdiffusion 
developed in Refs. \cite{Bao05,PRL} and presents 
thus a generalization
of this earlier model to the sub-ballistic case.  
Given
the lower integral limit $t_0=0$ in the GLE (instead of minus infinity), the exact 
reduction requires that the
$u_i(0)$ are independently Gaussian distributed with zero-mean and variance   
$\langle u_i(0)u_j(0)\rangle=k_BT\eta_i\delta_{ij}$ to ensure
the stationarity of the noise $\zeta(t)$ and FDR (\ref{fdt}) for all times.

Furthermore, on the time scale $t>\tau_h$, the diffusion 
becomes ballistic 
in our modeling. However, 
$\tau_h$ becomes {\it exponentially} larger with increasing $N$, which ensures, that 
the practical embedding dimension can be reasonably small ($D=15$ in simulations below). 
When speaking about asymptotic behavior below we yet
assume that $t<\tau_h$, but choose $\tau_h$ so, that it cannot be reached numerically.
This procedure is similar to the subdiffusive case \cite{G09}.

\section{Numerical Simulations}

Below, we investigate superdiffusion with $\alpha=1.5$ in 
a tilted washboard potential
of the form $V(x)=-V_0\cos(2\pi x/x_0)-Fx$.
For sake of convenience we transform the equations above into dimensionless
units by scaling time $t$ in units of $\tau_0=(\eta_\alpha/m)^{1/(\alpha-2)}$,
distance in $x_0$, energy (which applies to $V_0$, $k_BT$ and $Fx_0$) in 
$\Delta E=m(x_0/\tau_0)^2$
and $u_i$ in $mx_0/\tau_0^2$.
As a consequence of the time scaling, $\nu_0$ must be scaled in $1/\tau_0$, which implies, 
that $\eta_0/m$ is scaled in $1/\tau_0$ and $\eta_i/m$ in $1/\tau_0^2$.
 We integrated the dimensionless equations with a standard stochastic Euler algorithm
using a combination of Mersenne-Twister and Box-Muller algorithms to generate 
the Gaussian random numbers.
In each simulation an ensemble of $10^4$ particles was propagated with 
a time step $\Delta t=10^{-4}$ to achieve (weak) convergence of the 
ensemble averaged results. The end point of simulations is $t_{\rm max}=10^4$.
We used the friction kernel parameters of the approximation shown in fig.
\ref{fig1} and distributed the particle velocities initially thermally
with $\langle v(0)\rangle=0$ and $\langle v^2(0)\rangle=k_BT/m=v_T^2$. 
All particles were set initially to the position $x(0)=0$.

\subsection{Free superdiffusion}
\begin{figure}
\onefigure[width=.45\textwidth]{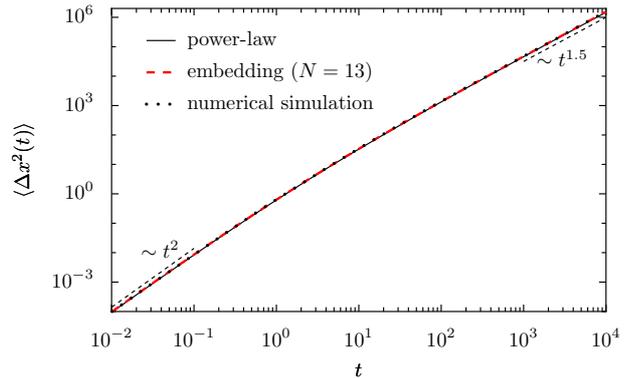} 
\caption{Comparison of the analytical FLE solution for
the mean-squared displacement in eq. (\ref{powex}) with the analytical results 
for the Markovian embedding and the corresponding numerical results, $\alpha=1.5$. }
\label{fig2} 
\end{figure}

We first test our method by comparison of the numerical results for free superdiffusion, 
i.e., $V_0=0$ and $F=0$, with 
the available analytical solution of FLE  \cite{Lutz}, 
\begin{align}\label{powex}
  \langle\Delta x^2(t)\rangle=2v_T^2t^2E_{2-\alpha,3}[-(t/\tau_0)^{2-\alpha}]\;.
\end{align}
Here, $E_{\alpha,\beta}(t)=\sum_{n=0}^\infty t^n/\Gamma(\alpha n+\beta)$
is the generalized Mittag-Leffler function. Furthermore, for  
$\eta(t)$ in eq. (\ref{fk}) the Laplace-transformed variance $\widetilde{ \sigma^2} (s)$
is a rational function which can also be inverted to the time domain.
The agreement in fig. \ref{fig2}
among the analytical FLE result, the analytical result for the Markovian embedding
and the simulation results
is indeed very good. Initially, the diffusion is always ballistic, 
$\langle\Delta x^2(t)\rangle\sim t^2$,  turning over
into the asymptotic behavior 
$\langle\Delta x^2(t)\rangle\sim t^{1.5}$.

\subsection{Superdiffusion in a tilted periodic potential}
Next, we study superdiffusion in a tilted periodic potential,
for which no analytical solution is available.
\begin{figure}
  \onefigure[width=.45\textwidth]{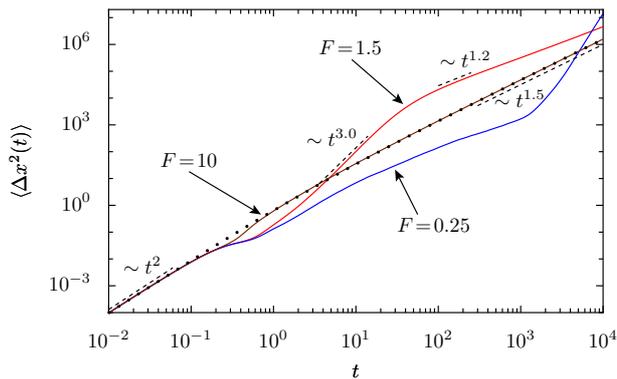}
  \caption{
 Position mean square displacement of particles in
a tilted washboard potential for several different values of 
the bias strength $F$ and the fixed potential amplitude
$V_0=1$ and the bath temperature $T=1$. The dotted line 
corresponds to the free case.
}
  \label{fig3}
\end{figure}
Similar to the free case, the superdiffusion starts 
again with ballistic diffusion and asymptotically again yields,  
$\langle\Delta x^2(t)\rangle\sim t^{1.5}$, cf. 
fig. \ref{fig3} for $F=10$.
However, on an intermediate time scale a {\it hyperdiffusive regime} is developed,
with $\langle\Delta x^2(t)\rangle\sim t^\lambda$, where $\lambda>2$.
Such an puzzling intermediate regime
(note that free superdiffusion cannot be faster than ballistic within the GLE description)
with a highly enhanced power-law dependence of the 
position variance
was also found recently for ballistic superdiffusion, see \cite{PRL}, and
seems to be a generic feature of driven superdiffusion in non-confining potentials.
Within the GLE description with a power law memory kernel
and a quite different numerical approach (not based on Markovian embedding),
 such a hyperdiffusion 
was revealed first 
in ref. \cite{Lu}. However, it was perceived there as an asymptotic regime probably 
because of insufficiently long time propagation. It is clear from fig. \ref{fig3}
that for a sufficiently small bias $F=0.25$ we also cannot arrive at the asymptotic regime
within two weeks of simulations. For larger bias strengths this is, however, possible.
The mean velocity of diffusing particles also grows with time and when the
corresponding mean kinetic energy becomes so large that the periodic potential 
ceases to play
a role, the diffusion attains asymptotically the regime of
free superdiffusion. However, on an intermediate time scale such a confined 
superdiffusion can be much faster than the free one. This surprising phenomenon
somewhat resembles giant acceleration of normal diffusion in washboard potentials 
\cite{giant}, but has a distinctly different origin \cite{PRL}.

The intermediate behavior of the position mean-squared displacement (MSD) 
is a consequence of
a transient heating of the particles from the bath temperature $T$ 
to the kinetic temperature $T_{\textrm{kin}}$,
which we define \cite{PRL} via the width of the velocity distribution
$k_BT_{\textrm{kin}}(t)/2\Doteq m\langle\Delta v^2(t)\rangle/2$,
$\Delta v(t)=v(t)-\langle  v(t)\rangle$. Such a kinetic temperature should
not be confused with the thermodynamic temperature. Nevertheless, it presents
a very helpful concept characterizing the kinetic energy of the 
disordered motion, when  the kinetic energy of the directed motion is subtracted
from the mean kinetic energy,
in accordance with the physical meaning of temperature.

\begin{figure}
  \onefigure[width=.45\textwidth]{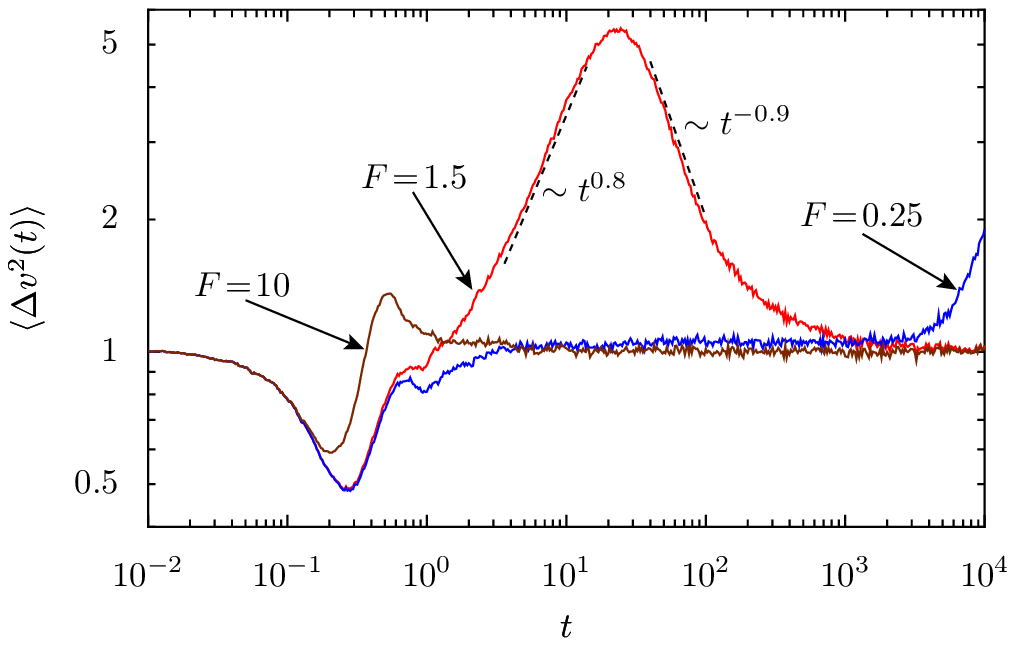}
  \caption{Mean square fluctuation of the particles velocity for the diffusion of 
  fig. \ref{fig3}.
  }
  \label{fig4}
\end{figure}
Fig. \ref{fig4} shows the time evolution of the velocity variance
for the diffusion in fig. \ref{fig3}. One can immediately see that 
the transient hyperdiffusion is related to the transient kinetic heating.
To explore this relation in more detail we notice that the position MSD 
equals the twice integrated VACF, 
i.e., $\langle\Delta x^2(t)\rangle=
2\int_0^t dt^{\prime}\int_0^{t^{\prime}} dt^{\prime\prime}
\langle\Delta v(t^\prime)\Delta v(t^{\prime\prime})\rangle$.
Introducing a normalized VACF
$K_v(t^{\prime\prime},t^{\prime})=
\langle\Delta v(t^\prime)\Delta v(t^{\prime\prime})
\rangle/\langle\Delta v^2(t^{\prime\prime})\rangle$
one can rewrite the integrand as
$\langle\Delta v(t^\prime)\Delta v(t^{\prime\prime})
\rangle=\langle\Delta v^2(t^{\prime\prime})\rangle K_v(t^{\prime\prime},t^\prime)$
and single out the evolution of the velocity variance. 
\begin{figure}
  \onefigure[width=.45\textwidth]{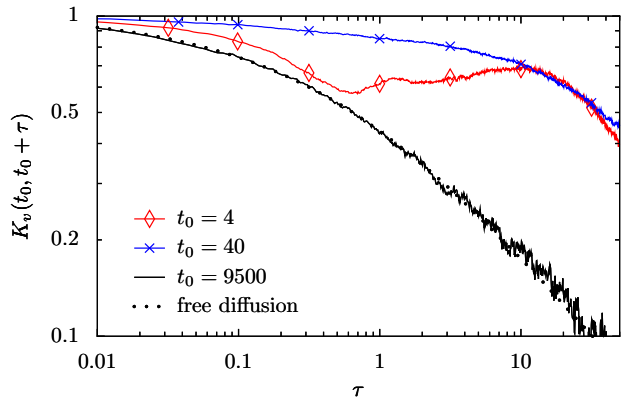}
  \caption{Normalized velocity autocorrelation function 
  for the free superdiffusion (dotted line, analytical result)
  and for the biased diffusion of fig. \ref{fig3} for $F=1.5$.
  One can see, that the latter decays much more slowly in the transient regime,
  for two values of $t_0$, which correspond to the start of the ascent of 
  the velocity variance in
  fig. \ref{fig4}, $t_0=4$, and the descent, $t_0=40$. In the
  asymptotic regime, $t_0=9500$, the VACF is again the same as
  in the free case. 
  }
  \label{fig5}
\end{figure}
If $K_v(t^\prime,t^\prime+\tau)$ would remain constant, or it would 
tend asymptotically to a constant with increasing $\tau$ (indicating
the breaking of ergodicity), then a power law increase of the
kinetic temperature with time, $T_{\rm kin}(t)\sim t^{\beta}$, would yield to a
hyperdiffusive law for the position MSD
$\langle\Delta x^2(t)\rangle\sim T_{\textrm{kin}}(t)t^2\sim t^{2+\beta}$.
This is what occurs for ballistic superdiffusion in tilted washboard potentials 
which is nonergodic \cite{PRL}.
There, the kinetic temperature increases until it saturates at a large 
final value \cite{PRL}. In contrast, the studied generalized Brownian
motion is ergodic. Its free motion VACF does decay to zero.
It also decays in the tilted washboard potentials, see in fig. \ref{fig5}
for the case $F=1.5$ in figs. \ref{fig3},\ref{fig4}.
However, this decay is enormously retarded in the transient regime, cf. 
fig. \ref{fig5}.
This retardation explains qualitatively the quasi-nonergodic 
origin of hyperdiffusion 
also in the present case. Indeed, if $K_v(t^\prime,t^\prime+\tau)$ were
a constant, then the hyperdiffusion power law exponent would be $\lambda=2+\beta\approx 2.8$
for the velocity variance growth depicted in fig. \ref{fig4}. This is not much different
from the observed value $\lambda\approx 3$ in fig. \ref{fig3}. However, 
very different
from the case of ballistic memory kernel \cite{PRL}, the kinetic temperature, associated with
the velocity variance, starts to decline towards the bath temperature, see in fig. \ref{fig4}.
This is because the motion is ergodic and after
the periodic potential ceases to play a role the velocity ACF has to decay to zero 
much faster (like in the free case). Since the dissipation is 
much stronger than in the ballistic case, the particles are not strictly accelerated 
by the bias $F$. 
 Their mean velocity grows rather sublinearly,
$\langle v(t)\rangle \sim F t^{\alpha-1}$ and one can  show
that for any $\alpha<2$ 
the mean kinetic energy of particles becomes negligible in the course of time 
as compared
with the work done by the biasing force $F$ on its increase.
This is very different to the 
ballistic case, where asymptotically a finite portion of work is used for the heating 
(see online Supplement of ref. \cite{PRL}), and leads eventually to 
the decline of the kinetic energy of disordered motion, 
cooling back to the bath temperature $T$.
 If our explanation of the transient hyperdiffusion in terms of a delayed decay of VACF
is correct, it should be able to explain another feature in fig. \ref{fig3}.
Namely, that the hyperdiffusive regime first turns over into a transiently 
decelerated superdiffusion with $\lambda\approx 1.2<\alpha$ 
(see for $F=1.5$), before $\lambda$ grows
back to $\alpha$. Indeed, after reaching the maximum in fig. \ref{fig4}
the velocity variance starts to decay in accordance with a power law $\beta\approx -0.9$.
Then, our reasoning with a strongly delayed VACF decay 
yields $\lambda=2+\beta\approx 1.1$ 
which is close to
the observed $\lambda\approx 1.2$ in fig. \ref{fig3}. 
This confirms that our line of reasoning
is consistent.

\section{Summary and conclusions}
In this work we proposed a general and simple method for Markovian 
embedding of a superdiffusive non-Markovian GLE dynamics and showed that it can
be used to approximate FLE dynamics over many time decades,
serving also as an independent approach to model superdiffusion.
We studied numerically such a superdiffusion with $\alpha=1.5$ in a tilted
washboard potential.
Concordant with our prior findings for the ballistic diffusion case \cite{PRL}, 
a hyperdiffusive transient regime was found.
We gave a simple physical explanation of this transient regime
as a kinetic heating effect in terms of the growing velocity variance
and a strongly delayed decay of the velocity autocorrelation function.
This is similar to the case of {\it nonergodic} 
ballistic superdiffusion in washboard potentials.
The transient can be very long and superdiffusion can become enormously 
accelerated, compared to the free superdiffusion, during this transient regime. 
However, very different from the ballistic case, the transient heating
is followed by a subsequent cooling back to the temperature of the thermal bath,
after the kinetic energy of the particles, which grows in time, 
exceeds much the potential energy. This cooling effect is due to the fact, that the motion
remains ergodic in the studied case 
and the friction is
sufficiently strong to take off the extra part of the kinetic energy,
which was built up during the transient heating regime. 
During this transient cooling regime the power law exponent of the diffusion
becomes {\it less} than the one of the free superdiffusion and gradually grows to the latter
in the course of time.  Then  the VACF coincides  asymptotically
with the one of the free motion. The influence of the periodic potential becomes forgotten,
very differently from the nonergodic ballistic case.  

In conclusion, we expect that our general methodology will be used in a number
of future applications of anomalous GLE diffusion. Moreover,  the 
surprising transient heating/cooling effects are expected to  
attract a further attention not only of theorists but
also of the experimental community.  

\acknowledgments
This work was supported by the German Excellence Initiative via the Nanosystems
Initiative Munich (NIM).

\end{document}